\documentclass[10pt,twoside]{article} 
\usepackage{graphicx}
\usepackage{verbatim}
\usepackage{framed}
\usepackage{hyperref}
\usepackage{listings}
\usepackage{amsmath}
\usepackage{morefloats}

\begin{document}

\title{Towards Efficient Computation of Functional Determinants in QM and QFT}
\author{Musa Maharramov \href{mailto:maharram@stanford.edu}{maharram@stanford.edu}} 
\date{\today}

\maketitle

\setcounter{secnumdepth}{2}

\section{Introduction}

Feynman Path Integrals (\cite{FPI}) offer a more intuitive alternative to the Schr\"{o}dinger approach in Quantum Mechanics. Green's functions for the Schr\"{o}dinger operator are computed as the limit of sums of exponents of the classical action along \emph{all} possible trajectories connecting two configurations on a lattice, as the lattice spacing tends to zero. Path integrals naturally reveal the classical trajectories as the limit of ``trajectory beams'' that contribute the most to the conditional probability. This naturally leads to asymptotics that use only \emph{some} of the trajectories -- e.g., those ``sufficiently close'' to the classic trajectories -- in the integral evaluation.

However, with the exception of a few special cases of simple Hamiltonians\footnote{e.g., a free particle, Harmonic Oscillator, a linear one-dimensional field}, Feynman path integrals are notorious for their computational complexity, even for reduced ``beams'' of trajectories. Computing path integrals using the \emph{classic trajectories only}, on the other hand, results in wrong Green's function amplitudes (although may still be useful for simple qualitative analysis of quantum phenomena as in the double-slit experiment \cite{BD}) However, such simplified Green's functions can be corrected by applying a normalizing ``pre-factor'' that can be shown to be the regularized \emph{operator determinant} for the \emph{equation in variations}. Therefore, our ability to compute functional determinants is key to successful application of path integrals to semi-classical approximation.
A natural question to ask in this context is why not use alternative semi-classical approximation techniques, such as WKB? (\cite{LL}) Methods based on asymptotic solution of the Schr\"{o}dinger equation represent the wave functions (or eigenmodes and energy levels if eigenvalue problem is being solved) as series of powers of $\hbar$. Specifically, the probability density is represented as  \[\psi(x,t)=A(x)\exp{\left(-\frac{ i S(x,t)}{\hbar}\right)}\] where \[A=A_0+\hbar A_1 + \hbar^2 A_2+\ldots\] and \[S=S_0+\hbar S_1 + \hbar^2 S_2+\ldots\].

However, it can be shown that for many cases of interest this asymptotic representation causes ``caustics'' -- i.e., multi-valued folding singularities -- where  the classical trajectories intersect (\cite{MF}). An elegant solution to the problem of caustics is delivered by Maslov's \emph{Canonical operator} (\cite{OM},\cite{MF}) that provides global-time asymptotic solution by ``splicing'' together \emph{local} asymptotics in $x$ and $p$-representations. The Canonical Operator effectively provides asymptotic solution on the phase-space as opposed to the traditional WKB that constructs asymptotics in $x$, thus on the configuration space. The disadvantage of the Canonical Operator is that, very much like the traditional WKB, it is essentially a device for avoiding caustics, and the splicing procedure -- while useful theoretically -- is problematic in practical applications of both Quantum Mechanics and wave propagation where short-wavelength approximations are routinely used.

Functional Integration, on the other hand, is an inherently ``global'' procedure that automatically takes care of ``widening'' or ``narrowing'' of the contributing beam of trajectories if \emph{all} the possible trajectories are included.
In this work we will review analytical and computational techniques for evaluating functional determinants for one and \emph{multi-dimensional} Hamiltonians. We will see that for the one-dimensional and radially-symmetric cases, operator determinants can be computed numerically using properties of the \emph{Fredholm} determinants of Storm-Liouville operators (\cite{QM4M}). We provide some numerical examples of computing operator determinants and cite references to applications in e.g. Quantum Field Theory. However, the case of arbitrary multi-dimensional Hamiltonians appears to be significantly more challenging, and apparently no computational technique exists that does require the explicit evaluation of at least some operator eigenvalues. We offer some tentative ideas for addressing the problem of multi-dimensional Hamiltonians, and discuss the associated computational challenges.

\section{Feynman Path Integral}

Solution to the Cauchy problem for the wave function describing motion of a single particle of mass $m$ in a potential field $V(x)$
\begin{equation}
i\hbar \frac{\partial \psi}{\partial t}\;=\;-\frac{\hbar^2}{2 m}\Delta \psi+V(x)\psi,
\label{eq:se}
\end{equation}
with the initial condition
\begin{equation*}
\psi(x,t)=\psi_{t_0}(x),\;t=t_0
\end{equation*}
is given by
\begin{equation}
\psi(x,t)\;=\;\int\limits_{\mathbf{R}^n}{K(x,t;y,t_0)\psi(y,t_0)dy}
\label{eq:prop}
\end{equation}
where the integration kernel, Green's function or \emph{propagator}
\begin{equation}
K(x,t;y,t')\;=\;\exp{\left[ -\frac{i(t-t')}{\hbar}H \right]}
\label{eq:kernel}
\end{equation}
has the meaning of the conditional probability of finding the particle at a point $x$ at time $t$ given that it was at the point $y$ at time $t'$, where $H$ is the \emph{Hamiltonian} from the right-hand side of (\ref{eq:se}). The operator exponent in (\ref{eq:kernel}) is difficult to compute for an arbitrary $V(x)$, as it requires the knowledge of eigenfunctions and energies. Evaluating the exponent of the Laplacian and the potential \emph{separately} is straightforward, but exponent of the sum of non-commuting operators is not equal to the product of individual exponents. However, commutator of two ``infinitesimal'' operators is an infinitesimal of a higher order, so we can use the following approximation
\begin{equation}
\exp{ -\frac{i\Delta t}{\hbar} \left[ \frac{p^2}{2m} +V(x)\right]}\approx \exp{\left[ -\frac{i\Delta t}{\hbar}  \frac{p^2}{2m} \right]} \exp{\left[ -\frac{i\Delta t}{\hbar} V(x)\right]}
\label{eq:small}
\end{equation}
for the propagator over a small time period $\Delta t$. Using the semigroup property of propagators and substituting (\ref{eq:small}) in (\ref{eq:prop}) with $\psi(x,t_0)=\delta(x)$, we obtain, after passing $\Delta t\to 0$:
\begin{align}
K(x,t;y,t')\;=\; & \lim_{N\to\infty}  \int\limits_{\mathbf{R}^{N-1}} \exp 
\left[\frac{i}{\hbar}\sum\limits_{j=0}^{N-1} p_j(x_{j+1}-x_j)-H(p_j,x_j)\Delta t \right]
\frac{dp_0}{2 \pi \hbar} \times \nonumber \\
& \prod_{j=1}^{N-1}\frac{dp_j dx_j}{2 \pi \hbar},
\label{eq:FPIphase}
\end{align}
where $H$ is the \emph{symbol} of the Hamiltonian and  we have substituted the free-particle kernel due to the first exponential in the right-hand side of (\ref{eq:small}). Note that (\ref{eq:FPIphase}) is the limit of summation over trajectories in the \emph{phase space} $(p,x)$ and the integrand is the Legandre transform of the Hamiltonian into the Lagrangian (\cite{MMCM}) times the time step. The equivalent \emph{configuration space} expression is
\begin{align}
K(x,t;y,t')\;=\; & \lim_{N\to\infty} \sqrt[n/2]{ \frac{m}{2 \pi i \hbar \Delta t}} \times \nonumber \\
 & \int\limits_{\mathbf{R}^{N-1}} \exp 
\left[\frac{i}{\hbar}\sum\limits_{j=0}^{N-1} \left(\frac{m}{2}\left(\frac{x_{j+1}-x_j}{\Delta t}\right)^2-V(x_j)\right)\Delta t \right]
\frac{dp_0}{2 \pi \hbar}\prod_{j=1}^{N-1}dx_j
\label{eq:FPIconf}
\end{align}
where the Fresnel integral formula was used for $p$-integrals. Note that if imaginary time is used, the procedure remains the same but the Fresnel integral is replaced with a Gaussian one. Formula (\ref{eq:FPIconf}) is interpreted as the summation of the exponentiated classical \emph{action} (\cite{MMCM}) times $\frac{i}{\hbar}$ over arbitrary discretized trajectories $x(t_j)=x_j$.

\section{Semi-classical Green's Function Approximation} 

Since we are interested in semi-classical approximations and would like to limit the summation in (\ref{eq:FPIconf}) to a \emph{narrow band of trajectories} around the classical trajectory, let us estimate the \emph{variation of the classical action due to perturbations of the classical trajectory}. Assuming that the perturbation is zero at $t$ and $t'$, replacing the sum in (\ref{eq:FPIconf}) with an integral from $t'$ to $t$ and using integration by parts we get:
\begin{align}
\sum\limits_{j=0}^{N-1} \left(\frac{m}{2}\left(\frac{x_{j+1}-x_j}{\Delta t}\right)^2-V(x_j)\right)\Delta t \approx & \nonumber \\
\int\limits_{t'}^{t}{\left(\frac{m}{2}|z'|^2-V(z)\right)dt}\;+\;\int\limits_{t'}^{t}{ \left(\frac{m}{2} |\Delta z'|^2-\frac{1}{2}\frac{\partial^2 V(z)}{\partial x^j \partial x^k} \Delta z^j \Delta z^k \right)dt}+O(|\Delta z|^2),
\label{eq:var}
\end{align}
where $z(t)$ is the classical trajectory and $\Delta z$ is the perturbation. Substituting (\ref{eq:var}) in (\ref{eq:FPIconf}) we see that \emph{the first order correction} to Green's function computed over the classic trajectories is given by the \emph{multiplicative factor}
\begin{align}
\lim_{N\to\infty} \int\limits_{\mathbf{R}^{N-1}} \exp 
\left[\frac{i}{2 \hbar}\int\limits_{t'}^{t}{ \left(m |\Delta z'|^2-\frac{\partial^2 V(z)}{\partial x^j \partial x^k} \Delta z^j \Delta z^k \right)dt}\right]
\prod_{j=1}^{N-1}d\Delta z_j.
\label{eq:Corr}
\end{align}
The exponent in (\ref{eq:Corr}) is a \emph{quadratic form} corresponding to the \emph{Hessian} of the Harmonic oscillator (``Jacobi'' matrix). Integral (\ref{eq:Corr}) is similar to the Fresnel integral (or a Gaussian integral if time is changed to imaginary) and is equal to
\begin{equation}
 \epsilon \frac{(2 \hbar \pi)^{n/2}}{ \sqrt{|\det J|}},
\label{eq:det}
\end{equation}
where $|\epsilon|=1$ and 
\begin{equation}
J=-m\frac{d^2}{dt^2} \mathbf{I} -\frac{\partial^2 V}{\partial x^j \partial x^k}.
\label{eq:jacobi}
\end{equation}
Note that operator Hessian (\ref{eq:jacobi}) maps vector-functions but in the 1D case this is a simple Storm-Liouville operator
\begin{equation}
J=-m\frac{d^2}{dt^2} -V''(x).
\label{eq:jacobi1d}
\end{equation}

\section{Regularized Operator Determinants and Operator Zeta Function}

Formula (\ref{eq:det}) was not proved but introduced by analogy with finite-dimensional quadratic forms. For arbitrary \emph{differential operators} (\ref{eq:jacobi}), (\ref{eq:jacobi1d}) we will have to \emph{define} operator determinant that would match the conventional determinant in the finite-dimensional case. One such definition utilizes \emph{Operator Zeta-function} (\cite{Dunne3}):
\begin{equation}
\zeta_J(\tau)\;=\;\sum\limits_{n=1}^{+\infty}\lambda_n^{-\tau},
\label{eq:zeta}
\end{equation}
where $\lambda_n\,n=1,2,\ldots$ is the discrete spectrum of $J$. Given (\ref{eq:zeta}), \emph{operator determinant} is \emph{defined} as
\begin{equation}
\det J\;=\;\exp { -\zeta'_J(0) }.
\label{eq:opdet}
\end{equation}
Note that definition (\ref{eq:opdet}) coincides with the finite-dimensional determinant if $J$ has a finite discrete spectrum and the summation in (\ref{eq:zeta}) is limited to non-zero eigenvalues. The Hessian operator for typical potentials may not have a simple discrete spectrum unless some boundary conditions are imposed on a finite domain (e.g., Dirichlet boundary conditions corresponding to the infinitely high potential barriers at the boundary of interest (\cite{LL}). We assume that at least for computational purposes such boundary conditions have been imposed, effectively cutting off far-field potentials.

If operator (\ref{eq:jacobi}) eigenvalues are known, then formula (\ref{eq:opdet}) can be used to compute the determinant after the operator zeta function is \emph{analytically continued to zero}. The last step is necessary because (\ref{eq:zeta}) does not define $\zeta_J$ at zero. The computation can be performed using contour integration (\cite{Kirsten}). However, the most difficult part -- computation of the operator spectrum -- is obviously not addressed by this procedure.

\section{Computation for one-dimensional and Radially-symmetric Hamiltonians}

In the simplest 1D case, operator determinant for (\ref{eq:jacobi1d}) can be computed using \emph{Gelfand-Yaglom} theorem (\cite{Gelfand},\cite{Dunne1},\cite{Dunne3}): 
\begin{framed}
\noindent
If Dirichlet boundary conditions are imposed on an interval $[a,b]$, then determinant of the operator (\ref{eq:jacobi1d}) acting on the corresponding Sobolev space $\overset{\circ}{H_1}$ is given by the $y(b)$, where $y(x)$ is the solution to the \emph{initial-value} problem
\begin{equation} 
Jy\;=\;0,\;y(a)=0,\;y'(a)=1.
\label{eq:ivp}
\end{equation} 
\end{framed}
Problem (\ref{eq:ivp}) can be easily solved using e.g. the Runge-Kutta numerical method (\cite{ASCHER}) for \emph{arbitrary} $V''(x)$. As a demonstration of this method, we will compute the values of the Riemann's zeta function (\cite{A}) at integer points 2 and 4\footnote{value at 1 is infinity}. Using the formula (\cite{Dunne1})
\begin{equation}
\log \frac{\det \left(J+\lambda^2\right)}{\det J}\;=\; \sum\limits_{k=1}^{\infty}\frac{(-1)^{k+1}}{k}\lambda^{2k}\zeta_J(k),
\label{eq:zetacalc}
\end{equation}
we compute the left-hand side for a set of values of $\lambda$, assuming that Dirichlet boundary conditions are imposed on $[0,\pi]$ and letting the potential be equal to zero. Eigenvalues of $J$ are then trivial to compute and are equal to $k^2,k=1,2,\ldots$, hence the operator zeta function is simply $\zeta_J(\tau)=\zeta(2\tau)$. Now fitting the tabulated values of the left hand side of (\ref{eq:zetacalc}) with a 4th order polynomial, we get $\zeta_J(1)=\zeta(2)\approx 1.6$ and $\zeta_J(2)=\zeta(4)\approx 1.1$ -- both results are in reasonable agreement with the exact values (\cite{A}). Of course, the accuracy of this algorithm is limited only by the number of points used in the fitting\footnote{we used just 10 values with built-in {\tt Matlab} polynomial fitting}, while the accuracy of the operator determinant evaluation is limited only by the Runge-Kutta time step.

The Gelfand-Yaglom proposition, although not applicable directly to multi-dimensional operators, is part of a similar technique developed for \emph{radially symmetric} operators (\cite{Dunne-Kirsten}). An application of the radially-symmetric operator determinants to fluctuation determinant for false vacuum decay is presented in \cite{Dunne3}. Note, however, that the same problem is solved in \cite{Dunne4} using WKB.

\section{Way forward for Multi-dimensional Hamiltonians}

For non-symmetric multi-dimensional Hessians (\ref{eq:jacobi}) it is natural to ask if the operator determinant can be computed using some iterative updating procedure based on e.g. one-dimensional operator projections. For example, in case of a two-dimensional potential pit, solving an initial-value problem similar to (\ref{eq:ivp}) along \emph{classic trajectories} connecting random boundary points would yield the operator determinants of one-dimensional operator projections. While these one-dimensional determinants are obviously connected with the determinant of $J$, it is unclear how to reconstruct the latter from the former.

A more traditional approach is based on 
\begin{itemize}
\item estimating operator eigenvalue asymptotics;
\item estimating, from the analytical extension of (\ref{eq:zeta}), the error in (\ref{eq:opdet}) due to the discarded highest eigenvalues\footnote{note that due to the analytical extension to zero and differentiation, this error may be larger than that of (\ref{eq:zeta})}, and estimating the required number of the \emph{smallest} eigenvalues required to achieve the desired accuracy;
\item computing the required number of eigenvalues using an iterative method, e.g. based on Lancsoz iterations (\cite{Matrix}). 
\end{itemize}
 
Note that for discretization grid sizes of $\approx 10^3$ in each dimension, the corresponding sparse numerical matrices for e.g. non-symmetric 3D potentials have dimensions of $10^9 \times 10^9$. Unless only a few initial eigenvalues are required and a quick convergence can be expected, this method may be impractical, especially in comparison with traditional perturbation and asymptotic techniques.

\bibliographystyle{plain}
\bibliography{MMFD}
\end{document}